\documentclass[aps,prd,nofootinbib,showkeys,preprint,floatfix]{revtex4-1}
  
\usepackage{graphicx,amssymb}    
\usepackage{epsfig}
\usepackage{graphics}     
\usepackage{psfrag}     
\usepackage{epsfig}\parskip 5pt plus 1pt
\usepackage{bbm}
\usepackage{amsmath}
\usepackage{amssymb}
\usepackage{amsfonts}
\usepackage{mathrsfs}
\usepackage{graphicx}
\usepackage{color}
\usepackage{graphicx,amssymb}    
\usepackage{epsfig}
\usepackage{graphics}     
\usepackage{psfrag}     
\usepackage{listings}
\usepackage{ulem}
\usepackage{slashed}

\def\to{\rightarrow}

\def\3mups{M_{\Upsilon(3S)}}

\def\lsim{\;\raise0.3ex\hbox{$<$\kern-0.75em\raise-1.1ex\hbox{$\sim$}}\;}
\def\gsim{\;\raise0.3ex\Curtin:2013fra,hbox{$>$\kern-0.75em\raise-1.1ex\hbox{$\sim$}}\;}
\def\beq{\begin{equation}}   \def\eeq{\end{equation}}
\def\ba{\begin{array}}       \def\ea{\end{array}}
\def\bea{\begin{eqnarray}}   \def\eea{\end{eqnarray}}
\def\nn{\nonumber}
\def\k{\kappa}
\def\l{\lambda}
\def\b{\beta}

\newcommand{\AddrWur}{%
Institut f\"ur Theoretische Physik und Astronomie,
Universit\"at W\"urzburg\\
Am Hubland,
97074 W\"urzburg, Germany}

\begin{document}  

\title{Discovery of Charged Higgs through $\gamma\gamma$ final states }

\author{Debottam Das} \email{debottam.phys@gmail.com}\affiliation{\AddrWur}

\author{Lukas Mitzka} \email{lmitzka@physik.uni-wuerzburg.de}\affiliation{\AddrWur}

\author{Werner Porod} \email{porod@physik.uni-wuerzburg.de}\affiliation{\AddrWur}

\begin{abstract} 
Extending the Higgs sector by an additional $SU(2)_L$ doublet Higgs boson
implies the  existence of a charged Higgs boson $H^+$. The LHC experiments
search for such particle focusing on it decays into
leptonic and quark decay final states, namely $\tau \nu$,$cs$ and $tb$. However,
if the Higgs sector if further extended, e.g.\ by a gauge singlet as in the 
NMSSM, the charged Higgs boson can also decay into a light scalar or
pseudoscalar Higgs boson which itself decays further into a two photon
final state. We present here scenarios where $H^+$ is produced in top-quark
decays with a sizable cross-section such the corresponding signal is
well above the Standard Model background at the 13 TeV run of Large Hadron 
Collider (LHC) with an integrated
 luminosity  100 fb$^{-1}$.
\end{abstract}   

\maketitle

\section{Introduction} 
\label{sec:intro}

The discovery of a scalar particle at the LHC which resembles strongly the Higgs particle
of the Standard Model (SM) with $m_H \sim 125$GeV \cite{Aad:2012tfa,Chatrchyan:2012ufa}
has been a great stride so far. Even though this particle shares many of the properties
of the SM Higgs boson, it could still be a member of an extended Higgs sector,
see e.g.\ \cite{Bechtle:2014ewa} and references therein. The search for the corresponding
additional particles as well as for deviations in the properties of the  Higgs boson 
(see e.g.\ \cite{Curtin:2013fra}) is one of the major tasks of the LHC experiments  \cite{LHCHiggs}. 

A particular well studied class of models are supersymmetric extensions of the SM. In its minimal version 
the Higgs sector is a two Higgs doublet model of type II. However, there are several other possibilities
where the simplest one is adding a gauge singlet Higgs field. 
An extended Higgs sector also implies non-standard production and decay possibilities, in particular
for the additional Higgs particles. In case of the Next to Minimal Supersymmetric Standard Model (NMSSM) 
\cite{Maniatis:2009re,Ellwanger:2009dp} a challenging task will be to find the additional states which 
resemble mainly the gauge singlet ones (see e.g.\ \cite{King:2014xwa}). As the direct production is strongly suppressed, one can use for 
example cascade decays of
supersymmetric particles or heavier Higgs bosons to produce them
\cite{Stal:2011cz,Kang:2013rj,Cerdeno:2013cz,Christensen:2013dra,Beskidt:2013gia,Chen:2013emb,Bhattacherjee:2013vga}. 
Similarly, pair production of the lighter Higgs bosons can be potentially a very interesting probe 
\cite{Papaefstathiou:2012qe,Cao:2013si,
Gouzevitch:2013qca,Nhung:2013lpa,Ellwanger:2013ova}. Additionally, the singlet
scalar could also open up new avenues to search for the charged Higgs scalar at the LHC e.g.\ via the cascade decays
of the top quark, $t \to H^+ b \to W^+ \Phi b \to W^+ b f \bar f$ \cite{Drees:1999sb} 
where $\Phi=H_1 (A_1)$ is the lightest (pseudo)scalar Higgs boson \cite{Drees:1999sb} and $f=b,\tau, \mu$ 
depending on the kinematical thresholds. A light pseudo-scalar 
$A_1$ decaying into $\tau^+\tau^-$ has been searched for by CDF \cite{Aaltonen:2011aj} and bounds
have been set for masses  of about 9~GeV. In the context of LHC, it has been shown recently that the 
aforementioned scenarios can easily be probed either with existing data or in the future runs 
\cite{Rathsman:2012dp,Dermisek:2012cn}. In addition the process $pp \to H_3 \to W^\pm H^\mp$ has 
been considered \cite{Dermisek:2013cxa} with the subsequent decay of $H^\mp$ into $H_1$, where
$H_3$ is the heaviest scalar Higgs boson.
 
In this letter we investigate to which extent
the charged Higgs boson can still be produced via top-quark decays with a subsequent
decay of the latter into a light Higgs boson ($H_1$ or $A_1$). We will focus on an intermediate
 mass range of the scalar and pseudoscalar of about 60-80 GeV which is potentially challenging since
the main decay modes are a pair of gluons and/or charm/bottom quark pair. However, there is the possibility
of an enhanced rate for $\gamma\gamma$ in this mass range as we will show below.
Final states resulting from $t\bar{t}$ production and containing two photons
have a rather small cross-sections in the SM and, thus, we find excellent prospects for the
discovery of new physics in the next run at the LHC for scenarios
where $H^\pm$ decays  dominantly into $W^+ H_1/A_1$.

In the next section we will briefly summarize the main features of the
Higgs sector of the NMSSM and in section \ref{sec:numerics} two
examples are presented. We will demonstrate
how a charged Higgs boson as well as a light Higgs boson can
be discovered at the LHC using  Monte Carlo studies. In section 
\ref{sec:conclusions} we will draw our conclusions.

\section {The Higgs sector of the NMSSM and some phenomenological aspects}
\label{sec:model}

In this section we briefly summarize some main features related to the NMSSM Higgs sector.
The superpotential of the NMSSM can be specified as
\beq\label{eq:2}
W_{NMSSM} = \lambda \hat S \hat H_u \hat H_d + \frac{\kappa}{3} \hat S^3 + W_{MSSM} \; 
\eeq
where $W_{MSSM}$ refer to the Yukawa interactions of the matter fields with the 
Higgs doublets  already present in the MSSM. The
vacuum expectation value (vev) $s$ of the real scalar component of $S$
generates an effective $\mu$-term
\beq\label{eq:1}
\mu_{eff}=\lambda s\; .
\eeq
Moreover, the
Lagrangian of the NMSSM contains trilinear and bilinear 
soft SUSY breaking terms related to the singlet Higgs sector: 
\beq\label{eq:2a}
 -{\cal L}_{NMSSM}^{Soft} = m_{S}^2 |S|^2 +\Bigl( \lambda A_\lambda\,
H_u \cdot H_d \,S +  \frac{1}{3} \kappa  A_\kappa\,  S^3  +
\mathrm{h.c.}\Bigl)\; +....;
\eeq
where, $...$ refers to the soft SUSY breaking terms already present in the MSSM. 
The complete Higgs sector consists of 
\begin{itemize}
\item 3 CP-even neutral Higgs bosons $H_i$, $i=1,2,3$;
\item 2 CP-odd neutral Higgs bosons $A_1$ and $A_2$;
\item One charged Higgs boson $H^\pm$.
\end{itemize}
where the neutral sectors are admixtures of doublet and singlet Higgs fields.

The $2 \times 2$ mass matrix for the CP-odd
Higgs bosons ${\cal M}_{P}^2$ has in the basis ($A_{MSSM}, S_I$)  the
elements
\bea
{\cal M}_{P,11}^2 & = & \frac{2\, \mu_\mathrm{eff}\,
(A_\lambda+\kappa s)}{\sin 2\b} \; , \nn\\
{\cal M}_{P,22}^2 & = & \l (A_\lambda/s+4\k ){v_u v_d} -3\k
A_\k\, s\; , \nn\\
{\cal M}_{P,12}^2 & = &\l (A_\l - 2\k s)\, v
\label{eq:3}
\eea
where $v_u$, $v_d$ denote the vevs of $H_u$, $H_d$, respectively, $v =
\sqrt{v_u^2 + v_d^2}$ and, as usual, $\tan\beta =
v_u/v_d$. The entry ${\cal M}_{P,11}^2$ would resemble
the mass of the MSSM-like CP-odd scalar. Note, that the
singlet like-state can be relatively light. 
We order the mass eigenstates according to $m_{A_1} \le m_{A_2}$
and apply this also to the scalar sector.

In the CP-even sector, three states $H_i(i=1,2,3)$ are admixtures of the 
real components $H_u$, $H_d$ and $S$. The state $H_{SM}$, which could be
either $H_1$ or $H_2$, 
with the nearly SM-like coupling to the electroweak gauge 
bosons has a mass \cite{Ellwanger:2011sk,Badziak:2013bda}
\begin{equation}\label{eq:4}
m_{{H_{SM}}}^2 = M_Z^2 \cos^22\beta + \l^2v^2\sin^2 2\beta 
+ \rm{rad corrs} + \Delta_{\rm {mix}} \quad.
\end{equation}
The term $\Delta_\mathrm{mix}$ represents the
singlet-doublet mixing 
\beq\label{eq:5}
\Delta_\mathrm{mix} \simeq \frac{4\l^2 s^2 v^2 (\l - \k \sin 2\beta)^2}
{{\overline{M}_{H_{SM}}^2 - M_{S}^2}} \;
\eeq
where $\overline{M}_{H_{SM}}^2$ is $m_{{H_{SM}}}^2$ 
without the mixing term and $M_{S}^2$ is the mass of singlet like Higgs boson. In scenarios as 
considered here where all Higgs states are light enough, one can still have significant mixing 
among all these states. If the singlet like state $M_{S}^2 $ is lighter/heavier
 than $ \overline{M}_{H_{SM}}^2$, then $\Delta_\mathrm{mix}$ can even produce 
significant positive/negative contributions to the lightest Higgs state.

The mass of the charged Higgs scalar is given by
\beq\label{eq:6}
{M}_{H^\pm}^2 = 
M_A^2 + v^2 \left(\frac{g_2^2}{2} - \l^2\right) . 
\eeq
Clearly, it decreases with increasing $\lambda$. 
We stress that even if this discussion of the masses is mainly at tree-level,
we have included the complete one-loop corrections to the Higgs masses
\cite{Degrassi:2009yq,Staub:2010ty,Graf:2012hh} and the dominant two-loop corrections
\cite{Degrassi:2009yq} in the numerical examples below.

The phenomenology of $H^+$ can differ significantly within the NMSSM compared
to the MSSM, as it can potentially decay into the $W^+H_1(A_1)$ even if its
mass is below the $t$-quark mass. The latter will decay further into
$f \bar{f}$, $gg$ and $\gamma\gamma$ pairs.
It turns out that small values 
of $\tan\beta$ are preferred as both $BR(t \to bH^\pm)$ and 
$BR(H^\pm \to W^\pm H_1(A_1))$ are enhanced in this case \cite{Dermisek:2012cn}.
As mentioned in the introduction we are particularly interested
in $H_1$ and/or $A_1$ in the mass range of 60--80 GeV where the lower bound
is given by the requirement that the SM-like Higgs boson should not
decay dominantly into two lighter Higgs bosons.
It has recently been shown 
that such a light $A_1$ can be tested with the existing LHC data 
\cite{Rathsman:2012dp} if it decays dominantly into $b\bar b$ with a
branching ratio of about 90\%. 
As we will show, the $\gamma\gamma$ channel can also be an interesting probe in this case. 
Similarly, for small $\tan\beta$ the lightest CP-even Higgs scalar $H_1$, which is mainly a 
singlet-like state, can dominantly decay 
into gluon pairs and/or charm quark pairs if the residual $H_u$ component is 
more important than the residual $H_d$ component. Both channels do not offer much 
prospects  at the LHC. 
However, the decay into two photons can be enhanced if
the chargino is light \cite{Ellwanger:2010nf,Ellwanger:2011aa,SchmidtHoberg:2012yy}
and in case of $H_1$ this can be further enhanced by a light $H^+$. Clearly,
such light states are also subject to flavour constraints as we will discuss below.

\section{Benchmark scenarios} 
\label{sec:numerics}

For the numerical evaluation we use {\tt SARAH} 
\cite{Staub:2008uz,Staub:2009bi,Staub:2010jh,Staub:2012pb,Staub:2013tta} 
to generate a NMSSM version of {\tt SPheno} 
\cite{Porod:2003um,Porod:2011nf} to compute the Higgs and the
SUSY particle spectrum, along with various couplings,
decay widths, and branching ratios. For the calculation of flavour observables we
use the package {\tt FlavorKit} \cite{Porod:2014xia}.
The spectrum is calculated including the complete one-loop corrections for
all masses of supersymmetric particles and Higgs bosons \cite{Degrassi:2009yq,Staub:2010ty} and as well
the dominant two-loop radiative corrections for Higgs bosons \cite{Degrassi:2009yq}. 

The numerical examples below we have taken $m_t=173.1$~GeV. Moreover, they
are compatible with the following constraints:
\begin{itemize}
\item Squark masses except for stops and sbottoms are assumed to be around $\sim$1.5~TeV 
to alleviate LHC constraints from direct SUSY searches 
 \cite{ATLAS1,Chatrchyan:2014lfa}.
For the same reason we assume the gluino mass $m_{\tilde g}$ to be larger than 1.6 TeV. 
In case of third generation squarks, the ATLAS 
\cite{ATLAS:2013pla,ATLAS:2013cma,TheATLAScollaboration:2013xha,TheATLAScollaboration:2013aia,TheATLAScollaboration:2013gha,Aad:2012tx} and CMS collaborations
\cite{Chatrchyan:2013xna,CMS-PAS-SUS-13-004} have obtained a
limit of up to $750$~GeV for $m_{\tilde t_1}$ assuming a 100\% branching ratio into either 
$\tilde t_1 \rightarrow t \chi_1^0$ or $\tilde t_1 \rightarrow b \chi^\pm_1$. However, it has been
shown that these bounds are relaxed if multiple final states are possible at the same time
\cite{Han:2013kga,Brooijmans:2014eja}.
As this is the case for our parameter choices below we take a lower bound of 600 GeV for $m_{\tilde t_1}$.

\item A SM-like Higgs boson with a mass in the range $M_{H_{SM}} = 123 - 128$~GeV.
For this we have fixed the trilinear soft susy breaking terms to: $T_{b,\tau} =A_{b,\tau}y_{b,\tau} = -1$~TeV and
$T_{t}=A_{t}y_{t} = -2.8$~TeV. Moreover, we check that the Higgs sector is consistent with existing data by using 
{\tt HiggsBounds-4.1.1} \cite{Bechtle:2008jh,Bechtle:2013wla}.

\item 
The first two generations of slepton masses are assumed to be around 200 GeV to have consistent spectra
with the muon anomalous magnetic moment constraint. However, for our considerations below it does not matter if they
are heavier.

\item It is quite well known that a light $H^\pm$ can lead to potentially
large contributions to flavor physics observables. The most constraining ones are  
$BR(b \to s \gamma) = (3.43 \pm 0.21 \pm 0.07 \pm 0.24^{th}) \times 10^{-4}$ 
\cite{Misiak:2006zs,Mahmoudi:2007gd,hfag}, $\Delta
M_{B_s} = 17.69 \pm 0.08 \pm 3.3^{th}$~ps$^{-1}$ \cite{Ball:2006xx,hfag}, $\Delta M_{B_d}
= 0.507 \pm 0.004 \pm 0.091^{th}$~ps$^{-1}$ \cite{Ball:2006xx,hfag} and $BR(B_s \to \mu^+ \mu^-) = (2.9 \pm 0.7 \pm 0.29^{th})10^{-9}$ \cite{Aaij:2013aka,Chatrchyan:2013bka,Mahmoudi:2012un} {\footnote{We considered 
2013 web updated results from the http://www.slac.stanford.edu/xorg/hfag/}}.
In the context of NMSSM, these constraints  were studied in detail in \cite{Domingo:2007dx}. 
In the region of the parameter space where $\tan\beta$
is small, the branching ratio $BR(B_s \to \mu^+ \mu^-)$ can easily be satisfied.
However, the other three constraints are rather restrictive and we get values
which are about 35-55\% enlarged compared to the experimental values. 
They can be brought to consistent values within the experimental and theoretical
uncertainties
if one allows for small non-minimal flavour violating structures in the soft-SUSY
breaking mass parameters as has been shown for example in \cite{Hurth:2009ke,Bruhnke:2010rh,Bartl:2011wq}
in the MSSM context with hardly an impact on the here discussed signatures.
The flavour mixing parameters impact on the mass of the SM-like Higgs boson
\cite{Heinemeyer:2004by}
but consistency between the Higgs mass constraint and the $b$-physics 
requirements can be achieved in a sizeable part of the parameter space
\cite{Bartl:2012tx}.
\end{itemize}
We do not consider dark matter constraints in this work. Though the
thermal relic abundance can be satisfied by tuning 
the values of $M_1$, $M_2$, $\mu$ and the slepton mass parameters, the limits from
direct detection experiments on  the dark matter  can be very stringent, thanks to the
substantial Higgsino component in lightest neutralino and lightness of all
Higgs states in our examples. 
It is well-known that tuning the strange quark content of the nucleon
\cite{Ellis:2008hf} and exploiting the astro-physics uncertainties,
the direct
detection limits can be relaxed by $\mathcal{O}(10)$ \cite{Das:2010kb}.
Moreover, one can easily extend the model to include $R$-sneutrinos which
could be the lightest SUSY particles. This can change the 
dark matter phenomenology significantly without affecting the discussion below, see e.g.\
\cite{Gopalakrishna:2006kr,Asaka:2006fs,Arina:2007tm,Thomas:2007bu,Cerdeno:2009dv,Cerdeno:2011qv}. 

\begin{figure}[ht!]
\begin{center}
\includegraphics[scale=1]{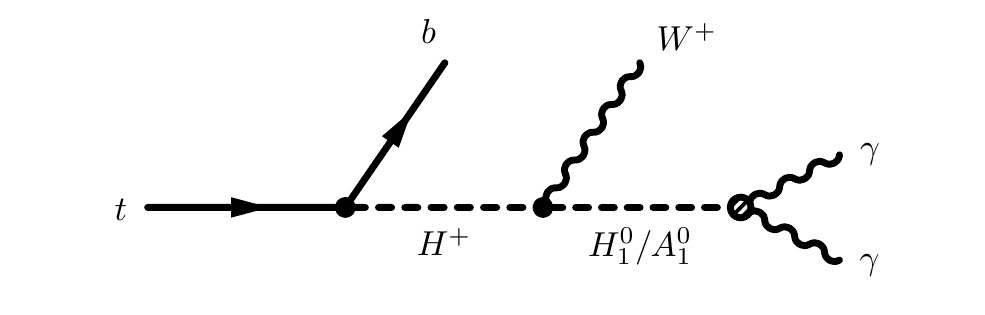}
\end{center}
\caption{Feynman diagram depicting cascade production of two photons from top decay via Higgs bosons.}
\label{fig:topDecayDiagramGammaGamma}
\end{figure}

In the table \ref{tabchg} we present two benchmark points BMP-A and BMP-B where
the charged Higgs boson is lighter than the top quark. 
In both cases we consider $t$-quark pair production
where one of $t$-quarks decays as usual into $W b$ whereas the second one
decays into $H^+ b \to \Phi W b \to \gamma \gamma  W b$ as depicted in Fig.~\ref{fig:topDecayDiagramGammaGamma}.
Here $\Phi$ is either $H_1$ (scenario BMP-A) or $A_1$ (scenario BMP-B). 
We focus on the $\gamma \gamma$ decay mode of $\Phi$ due to its clean signature at the LHC. 
Before continuing we note that in the first case the decay $\Phi \to b\bar{b}$
is suppressed as $H_1$ is mainly a gauge singlet
with a still sizeable $H_u$ component which not only gives the relatively large branching
ratio into $\gamma \gamma$ but also large branching ratios into $c\bar{c}$ and $gg$.

\begin{table}[htp]
\begin{center}
\begin{tabular}{|c|cc||c|cc|} \hline \hline
parameter      & BMP-A & BMP-B  & Branching ratios   & BMP-A &  BMP-B   \\ \hline \hline
$\tan\beta$     & 1.68  &1.45     & $Br(t\rightarrow bH^+)$    
& $3.3\times 10^{-3}$   & $1.8\times 10^{-2}$  \\
$\kappa$         & 0.596 & 0.94   & $Br(H^+\rightarrow W^+H_1)$ & 0.68  & 0     \\     
$\lambda$       & 0.596 &0.62 & $Br(H^+\rightarrow W^+A_1)$ 
& 0 & 0.86   \\
$\mu_{eff}$      & 131.5    &143.7   &   $Br(H_1\rightarrow \gamma \gamma)$ 
& $6.0\times 10^{-3}$     &~$1.76\times 10^{-3}$ \\
$m_{H_{1}^0}$      &73.8  & 126.8   & $Br(A_1\rightarrow \gamma \gamma)$  
&$2.6\times 10^{-5}$ &~$1.0\times 10^{-4}$ \\
$m_{H_{2}^0}$      & 126.7  & 172.0   & $Br(H_1\rightarrow b\bar b)$   &~ 0.24    &~0.72 \\
 $m_{H_{3}^0}$      & 192.6  & 364.1   & $Br(A_1\rightarrow b\bar b)$ &~ 0.11    &~0.86  \\ 
$m_{A_{1}^0}$       & 155.4& 66.6  &$Br(H_1\rightarrow c\bar c)$&~0.32&~0    \\
$m_{A_{2}^0}$       & 428.4& 402.0  &$Br(A_1\rightarrow c\bar c)$&~0&~0  \\
$m_{H^{\pm}}$      & 161.9 & 149.4  &$Br(H_1\rightarrow g~g)$&0.4&0.02 \\
$m_{\tilde t_1}$       & 645.4& 656.2 &$Br(A_1\rightarrow g~g)$&~0&0.02   \\
$m_{\tilde b_1}$      & 883.5 & 879.0 &&&  \\
\hline \hline
\end{tabular}
\end{center}
\caption{Relevant masses and branching fractions of the benchmark points used for our simulation. 
All masses are in GeV.}
\label{tabchg}
\end{table}

\begin{table}[h]
\begin{center}
\begin{tabular}{| c | c | c |c |c |c |c | }\hline
$p_{T_{min}}(j) $ & $ p_{T_{min}}(\gamma) $  & $p_{T_{min}}(\ell) $ & $|\eta|_{max}(j)$ & $|\eta|_{max}(\gamma)$ & $|\eta|_{max}(\ell)$ & $ \Delta R $   \\
\hline
20 GeV & 10 GeV  & 10 GeV & 5 & 2.5 & 2.5 & 0.4   \\
\hline
\end{tabular}
\end{center}
\caption{The default cut setup in MadGraph 5.1.5.13 where $p_T$ is the transverse momentum, $\eta$ the pseudorapidity 
and $\Delta R$ for the angular distance between two objects comprised of leptons, jets and photons.}
\label{tab:defaultMadGraph}
\end{table}

We consider first the cross-section for the signal
\begin{align}
\sigma^{\Phi}_{2b+2W+2\gamma} = 2 \times\sigma(pp \rightarrow t \bar t) \times
& Br(t \rightarrow bH^+) \times Br(\bar t \rightarrow \bar b W^-) \nonumber \\ \times 
& Br(H^+\rightarrow W^+\Phi)\times Br(\Phi\rightarrow \gamma \gamma)
\end{align}
with $\Phi = H_1,A_1$. For a qualititative understanding we calculate the effective signal events, as defined
by $\mathcal {S}^{H_1(A_1)} \equiv \sigma^{H_1(A_1)}_{2b+2W+2\gamma} \times \mathcal {L}$, where $\mathcal {L}$ represents
the integrated luminosity for present or future LHC runs. The results are 
shown in Tab.~\ref{tab:events} which have been calculated with  MadGraph 5.1.5.13 \cite{Alwall:2011uj}
using its default cut setup as shown in Tab.~\ref{tab:defaultMadGraph} and with the CTEQ6L1 PDF set.
These numbers have of course to be compared with the SM background processes. The dominant one is obviously
the irreducible one: $pp \to WWb\bar{b} + \gamma \gamma$. In addition there are two more due to the fact
that $b$-jet coming from the $t\to H^+ b$ is rather soft due to the small mass difference $m_t-m_{H^+}$:
(1) For $10$~GeV$<p_T(b)<25$~GeV, where $ p_T(b)$ is the transverse momentum of the
$b$-jet, the $b$-jet can be reconstructed as a jet but
its flavour cannot be identified anymore. The corresponding background is $WWbj + \gamma \gamma$.  
(2) For $p_T(b)< 10$ GeV, the jet cannot be reconstructed at all and, thus, we take also $WWb + \gamma \gamma$
as a background into account. The cross-sections for all three background reactions are given in 
Tab.~\ref{tab:bck1}
\begin{table}[t]
\begin{center}
\begin{tabular}{| c | c | c|c|}\hline
  CM energy & $\mathcal {L} (fb^{-1})$
	& 
BMP-A ($\mathcal {S}^{H_1}_{2b+2W+2\gamma}$) &BMP-B ($\mathcal {S}^{A_1}_{2b+2W+2\gamma}$)  \\
    \hline
8~TeV  &20    & 121  & 14   \\
13~TeV &100 & 1987  & 228\\
14~TeV &100 & 2351  & 270 \\
\hline
\end{tabular}
\end{center}
\caption{Total signal events $\mathcal {S}^{H_1(A_1)}_{2b+2W+2\gamma}$ using leading order $\sigma(p p \to t \bar t)$.}
\label{tab:events}
\end{table}
Comparing both tables we see that in the first scenario already a trivial counting of the events without
any further cuts gives a clear indication that there is physics beyond the SM involved as the numbers
for the signal and the background are of the same size.

\begin{table}[t]
\begin{center}
\begin{tabular}{| c | c | c | c| }\hline
 Background events & 8 TeV & 13 TeV & 14 TeV  \\
    \hline
$pp\to W^+ W^- b \bar{b} \gamma \gamma$ & 181  & 2859 & 3353 \\
$pp\to W^+ W^- \overset{(-)}{b} j \gamma \gamma$ & 5  & 146 & 261 \\
$pp\to W^+ W^-   \overset{(-)}{b} \gamma \gamma$ & 9 & 194 & 240 \\
\hline
\end{tabular}
\end{center}
\caption{Number of the different  SM background events without the subsequent decays of the $W$ bosons
using the default cuts given in Tab.~\ref{tab:defaultMadGraph}. 
}
\label{tab:bck1}
\end{table}

In case of scenarios like  BMP-B one needs of course further cuts to extract the corresponding signal.
We have  performed  Monte Carlo studies for both scenarios at the parton level for this channel at the LHC with
13 TeV c.m.s.~energy and assuming an integrated luminosity of $100$ fb$^{-1}$.
For the signal process with one $t$ decaying as depicted in Fig.~\ref{fig:topDecayDiagramGammaGamma} we use an 
implementation of the NMSSM to MadGraph that has been obtained from SARAH via the SUSY toolbox \cite{Staub:2011dp}. 
The $H^0_1/A^0_1\to \gamma\gamma$ process is performed with Pythia \cite{Sjostrand:2006za}. 
The background processes are generated with MadGraph.
We have generated $10^4$ events for the signal of the aforementioned benchmark points and its background processes from 
Tab.~\ref{tab:bck1} assuming that one of the $W$'s decays hadronically and the other one leptonically to $e$ or $\mu$.
We weight the generated events according to the cross-sections. 

\begin{figure}[ht!]
\begin{center}
\includegraphics[scale=.4,angle=270]{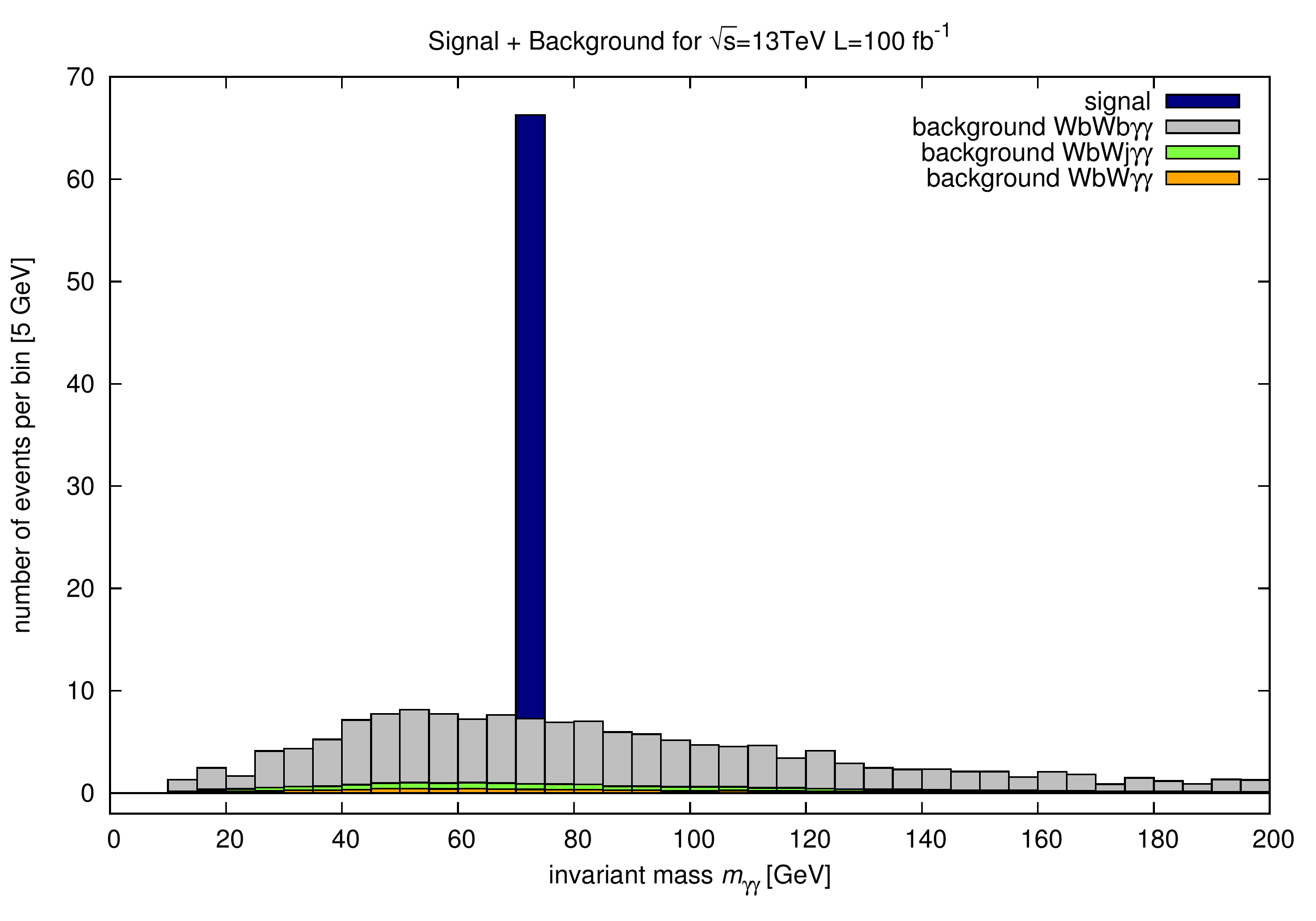}
\end{center}
\caption{Invariant mass distribution of the photon pair for BMP-A of signal (blue) and background 
contributions stemming from $p p \to W^+ b W^- \bar{b} \gamma \gamma $ (grey), 
from $p p \to W^+ \overset{(-)}{b} W^- j \gamma \gamma $ (green) and from $p p \to W^+ \overset{(-)}{b} W^- \gamma \gamma$ (orange)  
with one W decaying hadronically and the other one leptonically.}
\label{siggg}
\end{figure}

\begin{figure}[ht!]
\begin{center}
\includegraphics[scale=.4,angle=270]{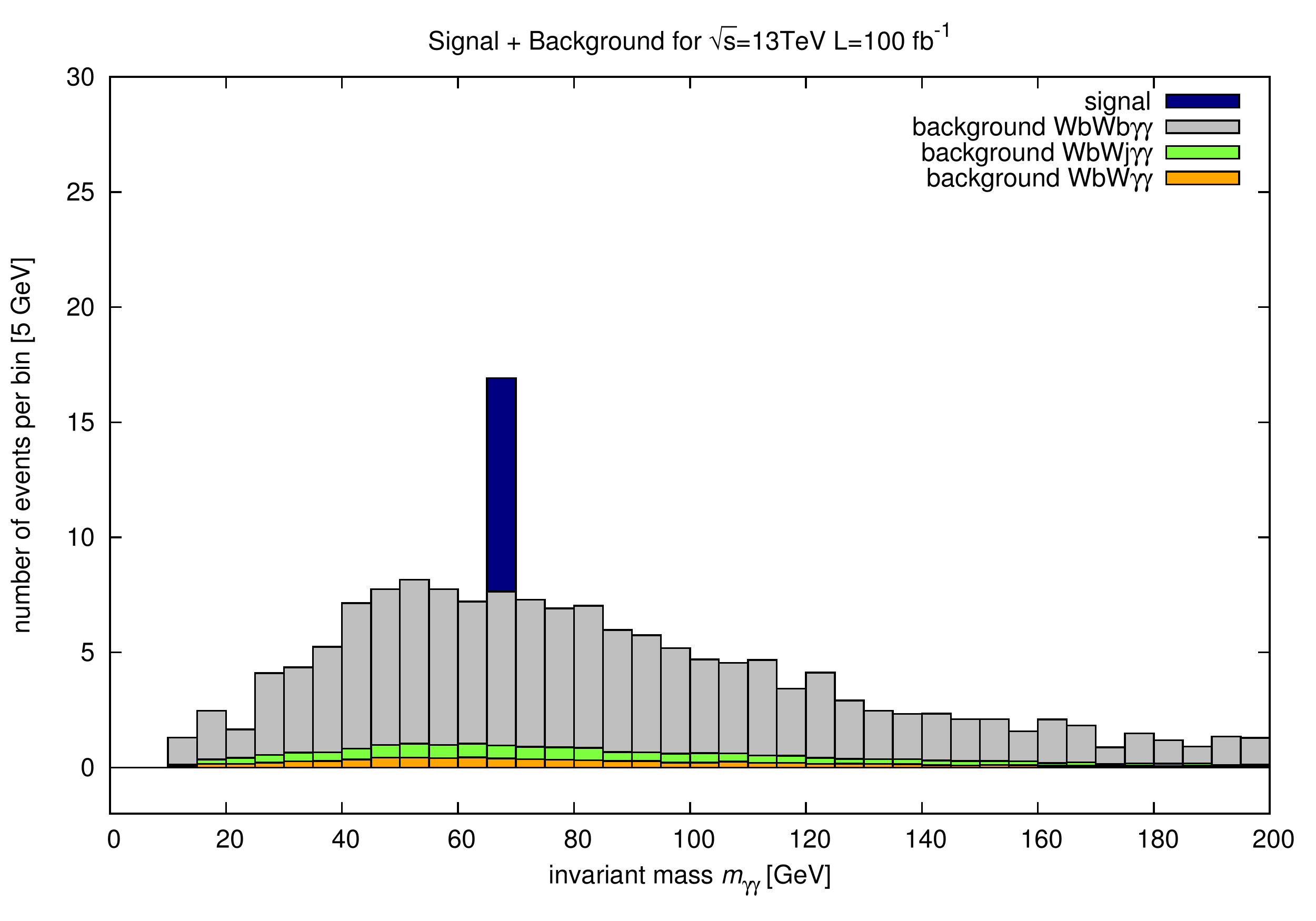}
\end{center}
\caption{
Same as in Fig.~\ref{siggg}, but for BMP-B.}
\label{sigggODD}
\end{figure}

In Fig.~\ref{siggg} and  Fig.~\ref{sigggODD} we plot the distribution of the invariant mass $m_{\gamma\gamma}$ of the 2-photon system 
for the signal and background processes using the cuts of Tab.~\ref{tab:PlotCuts}.
We check that all identifiable objects, i.e. everything except the non-identifiable soft $b$ jets, are well separated 
from each other with $\Delta R > 0.4$ and have $|\eta| < 2.5$.
We demand that the two hardest jets with $p_T > $ 20~GeV have an invariant mass $m_{jj}$ within a window  of 
$\Delta^{jj}_{W}=$20~GeV around the $W$ mass. Then we require two photons with $p_T >$ 20~GeV and an invariant mass $m_{\gamma\gamma}$ of 
at least 10 GeV. Finally we demand one lepton with $p_T > $ 10~GeV and that the hardest $b$ has $p_T > 40$ GeV. 
As has to be expected from the above considerations one sees a clear signal peak over the background
for the scenario  BMP-A in Fig.~\ref{siggg}. In case of  BMP-B one sees in Fig.~\ref{sigggODD} that at least
at the parton-level one has a clear signal over the background. However, in this case a full detector study
will be necessary to check if this still holds under more realistic assumptions.

\begin{table}[t]
\begin{center}
\begin{tabular}{| c | c | c | c |c |c|}\hline
$p_{T}(j) $ & $ m_{\gamma\gamma} $  & $p_T$ of hardest $b$ & $p_T(\gamma)$ & $\Delta^{jj}_{W}$ & $\Delta R_{min}$ \\
\hline
20 GeV & 10 GeV  & $>$ 40 GeV & 20 GeV & 20 GeV & 0.4  \\
\hline
\end{tabular}
\end{center}
\caption{
Cuts used for our simulation. The minimum $p_T$ for jets has to be larger than 20 GeV for the two hardest
jets and larger than 10 GeV for the softer ones.  At least one of the two $b$ quarks must have a $p_T>40$ GeV. 
The invariant mass of the two hardest non-btagged jets is required to fulfil $|m_{jj}-m_W|<\Delta^{jj}_W$.
The invariant mass of the two photons $m_{\gamma\gamma}$  has to be larger than 10 GeV.
The photons need to have a $p_T > 20$ GeV. }
\label{tab:PlotCuts}
\end{table}

Last but not the least we want to stress, that all results have been obtained so far using tree-level
cross-sections. However, it is well known that the $t\bar{t}$ production cross-sections receives large
 QCD corrections. Using the online-program available at ref.~\cite{ttXs} we have calculated
the top pair production cross-section $\sigma(pp \rightarrow t \bar t)$ including NLO+NNLL corrections
\cite{Cacciari:2011hy}. Here we have taken 
for $m_t=173.1$~GeV and the PDF-set MSTW2008nnlo68cl \cite{Martin:2009iq}. Compared
to the tree-level results used above we obtain a K-factor of 1.7, 1.6 and 1.6 for LHC
7, 13 and 14 TeV c.m.s.\ energy, respectively. In case that the background could
be rescaled by a similar factor, this would imply an improvement of the signal over square
root background ratio of about 30\%.
\begin{table}[h]
\begin{center}
\begin{tabular}{|c|c|c|c|} \hline
Channel:      & 8 TeV & 13 TeV  & 14 TeV  \\\hline
$\sigma$ [fb] & 228$\times 10^3$ &746 $\times 10^3$ & 882 $\times 10^3$ \\
$\sigma_{LO-MG5}$ [fb]& 135 $\times 10^3$ & 463 $\times 10^3$ & 555 $\times 10^3$ 
\\\hline
\end{tabular}
\end{center}
\caption{Signal cross-sections for $\sigma(pp \rightarrow t \bar t)$ given in NLO+NNLL accuracy and 
at leading order according to MadGraph. 
}
\label{tab:tt}
\end{table}
\section{Conclusions}
\label{sec:conclusions}
Within the framework of the NMSSM the charged Higgs boson can dominantly decay
into the  lightest Higgs scalar $\Phi$~($\Phi=H_1,A_1$) through 
$H^\pm \to W^\pm \Phi$. Subsequently, the lightest Higgs scalar 
can decay into $\gamma \gamma$ which
leads to a novel channel for the discovery of $H^\pm$ at the LHC. We 
have demonstrated this for two scenarios with $m_{\Phi} \in 60-80$ GeV.
 Our simulations at the parton-level
delineate the clear excess of signal events over the backgrounds in the
considered mass range of $m_{\Phi}$ which can easily be seen at the next runs
of LHC. This will endorse the presence of a light Higgs scalar and a light 
charged Higgs boson of an extended Higgs sector in a supersymmetric framework.
Thus LHC collaborations should expand their search strategy to include 
the di-photon search channel for a light charged Higgs scalar to account for
this possibility. Last but not the least we note that the existence of such
a light charged Higgs boson necessitates a non-trivial flavour structure
in the squark sector to obtain consistency with the existing low energy data.

\subsection*{Acknowledgments}
We thank G.~Siragusa for discussions on b-tagging  at the ATLAS
experiment.
This work has been supported by DFG, project no.\
PO-1337/3-1 and by DFG research training group GRK 1147.
LM acknowledges  support from the Elitenetzwerk Bayern.
\vfill

\bibliography{CHG_NMSSM.bbl}

\bibliographystyle{h-physrev5}

\end{document}